\documentclass[pre,twocolumn,amsmath,amssymb]{revtex4-1}

\usepackage{graphicx} 
\usepackage{enumitem}
\usepackage{float}
\usepackage{physics}
\usepackage{amsmath}
\usepackage{balance}
\usepackage{color}
\usepackage{lastpage}
\usepackage{lineno}
\usepackage{dcolumn}

\usepackage[colorlinks=True]{hyperref}  
\hypersetup{
 	colorlinks=true,  
 	linkcolor=blue,  
 	citecolor=blue, 
 	filecolor=magenta,   
 	urlcolor=blue         
 }
 
\makeatletter
\def\maketitle{
\@author@finish
\title@column\titleblock@produce
\suppressfloats[t]}
\makeatother

\makeatletter
\newcommand\footnoteref[1]{\protected@xdef\@thefnmark{\ref{#1}}\@footnotemark}
\makeatother

\makeatletter
    \def\balanceissued{unbalanced}
    \let\oldbibitem\bibitem
    \def\bibitem{%
        \ifnum\thepage=6%
            \expandafter\ifx\expandafter\relax\balanceissued\relax\else%
                \balance%
                \gdef\balanceissued{\relax}\fi%
            \else\fi%
        \oldbibitem}
\makeatother

\def\etl{$et~al.$~}

\begin{document}
\title{Finite-time scaling for kinetic rough interfaces}

\author{Rahul Chhimpa}
\affiliation{Department of Physics, Institute of Science,  Banaras Hindu University, Varanasi 221 005, India}

\author{Avinash Chand Yadav\footnote{jnu.avinash@gmail.com}}
\affiliation{Department of Physics, Institute of Science,  Banaras Hindu University, Varanasi 221 005, India}

\date{\today}


\begin{abstract}
We consider discrete models of kinetic rough interfaces that exhibit space-time scale-invariance in height-height correlation. A generic scaling theory implies that the dynamical structure factor of the height profile can uniquely characterize the underlying dynamics. We provide a finite-time scaling that systematically allows an estimation of the critical exponents and the scaling functions, eventually establishing the universality class accurately. As an illustration, we investigate a class of self-organized interface models in random media with extremal dynamics. The isotropic version shows a faceted pattern and belongs to the same universality class (as shown numerically) as the Sneppen (model A). We also introduce an anisotropic version of the Sneppen (model A) and suggest that the model belongs to the universality class of the tensionless one-dimensional Kardar-Parisi-Zhang equation.
\end{abstract}

\maketitle
The phenomenon of \emph{kinetic surface roughening} (a dynamically growing rough surface or interface) occurs in diverse contexts, and it has been a topic of much interest, particularly in non-equilibrium statistical physics, in advancing theoretical understanding~\cite{Zhang_1995, Stanley_1995, Tauber_2014, Marchiori_2014}. Typical instances include fluid flow in porous media~\cite{Soriano_2002}, the spreading of fracture cracks~\cite{Schmittbuhl_1998, Ponson_2006}, and fungal growth~\cite{Jensen_1998}. In condensed matter physics, the study of the thin-film growth formed by particle deposition processes (for example, molecular-beam epitaxy~\cite{Sarma_1991, Lai_1991, Kim_1994}) seems important technologically.

Strikingly, many systems of kinetic surface roughening exhibit scaling features. Determining the universality class of the model has been a crucial aspect. A set of independent critical exponents characterizing the scaling properties of the rough surface determines the universality class. The most familiar classes are random deposition, Edwards-Wilkinson (EW)~\cite{Edwards_1982}, and Kardar-Parisi-Zhang (KPZ)~\cite{Kardar_1986, Sasamoto_2010, Calabrese_2011, Takeuchi_2018, Cuerno_2023}. Several discrete surface roughening models have been introduced and examined in the past to uncover the underlying mechanisms. Random deposition with surface relaxation or growth preferred at local minima~\cite{Family_1986, Kwak_2019} represents a discrete model of the EW class, while several models (Eden~\cite{Eden_1958}, ballistic deposition~\cite{Family_1985}, and restricted solid on solid~\cite{Kim_1989}) belong to the KPZ class. The turbulent liquid crystal~\cite{Takeuchi_2010, Takeuchi_2011} also falls into the KPZ class.

While surface roughening remains one aspect, several other properties have been of concern. For example, the distribution of width~\cite{Foltin_1994}, the maximal height~\cite{Shapir_2001}, the density of extrema~\cite{Toroczkai_2000}, the cycling effects~\cite{Shapir_2000}, and the maximal spatial persistence~\cite{Xun_2019}.

Let $h(x,t)$ be the height profile of a fluctuating interface on a one-dimensional substrate with $1\le x\le L$. The commonly used characterization of the height profile is the global interface width  \begin{equation}
w(t, L)  = \langle \overline{ [h(x,t)- \bar{h}(t)]^2} \rangle^{1/2}.
\label{eq_w1}
\end{equation}
The overline in Eq.~(\ref{eq_w1}) represents the average over all sites $x$, and the angular brackets $\langle \cdot \rangle$ denote the ensemble average over different realizations. For the scale-invariant rough interface, the global interface width exhibits the Family-Vicsek dynamic scaling ansatz~\cite{Family_1985}
\begin{equation}
w(t, L) = t^{\chi/z}f(L/\xi(t)).
\label{eq_w2}
\end{equation}
The correlation length varies as $\xi(t) \sim t^{1/z}$, where $z$ denotes the \emph{dynamical exponent}.
The scaling function $f(u)$ in Eq.~(\ref{eq_w2}) assumes a form
\begin{equation}
f(u) \sim \begin{cases} u^{\chi}, ~~~~~~~~~~~~~~~~~u\ll 1,\\ {\rm constant}, ~~~~~~~~~u\gg 1,\\ \end{cases}
\label{eq_w3}
\end{equation}
where $\chi$ is the \emph{roughness exponent} that characterizes the stationary regime $\xi(t)\gg L$. The \emph{growth exponent} $\beta = \chi/z$ describes the short-time behavior of the interface.

In many growth models, it has been found that while the local width (and height-height correlation) behave similarly to Eq.~(\ref{eq_w2}) as $w(t, l) = t^{\beta}f_l(l/\xi(t))$, the scaling function differs from Eq.~(\ref{eq_w3}) as
\begin{equation}
f_l(u) \sim \begin{cases} u^{\chi_{\rm loc}}, ~~~~~~~~~~~~~~u\ll 1,\\ {\rm constant}, ~~~~~~~~~u\gg 1.\\ \end{cases}
\label{eq_lw1}
\end{equation}
In Eq.~(\ref{eq_lw1}), the local roughness exponent $\chi_{\rm loc}$ is an independent exponent~\cite{Lopez_1999, Lopez_2000}.
This intriguing feature is the so-called \emph{anomalous roughing} and has been of much interest~\cite{Lopez_1996, Lopez_1997, Lopez_2005, Szendro_2007}. L\'opez suggested that the anomalous features emerge from the nontrivial dynamics of the mean square local slope $\langle \overline{(\nabla h)^2}\rangle \sim t^{2\kappa}$ with $\chi_{\rm loc} = \chi -z\kappa$~\cite{Lopez_1999}.
Recent studies have shown anomalous behavior in a class of kinetic rough interfaces externally driven by long-time correlated noise~\cite{Lopez_2019, Lopez_2020, Lopez_2021}.

Ramasco \etl \cite{Lopez_2000} observed that a \emph{generic scaling theory} for the structure factor can reveal the unique dynamical features, including the anomalous feature. However, to get data collapse for the dynamical structure factor, the exponents $\chi$ and $z$ need to be known (discussed below). The scaling analysis of the global and local interface widths can provide an estimate of the two exponents and a sign of the existence of anomalous feature, respectively. Alternatively, on a double logarithmic scale, the envelope and individual curve slopes for the dynamical structure factor can provide an approximate estimate of the roughness exponent $\chi$ and the \emph{spectral roughness exponent} $\chi_s$, respectively~\cite{Kolton_2019}. 
$\chi_s$ is a remarkable characteristic for understanding the broad subclass of the underlying process (discussed below). However, a precise estimate of the spectral roughness exponent remains missing. Moreover, a \emph{systematic analysis} of the scaling feature of the dynamical structure factor seems lacking. Ramasco \etl\cite{Lopez_2000} examined the Sneppen (model A)~\cite{Sneppen_1992}, a self-organized interface depenning in random media, showing faceted patterns ($\chi_s > \chi$) with $\chi_s = 1.35$. Our analysis, well supported numerically, reveals that the precise value of the exponent is $\chi_s = 3/2$.

In this paper, we provide \emph{finite-time scaling} (FTS), a systematic approach for the scaling analysis of the dynamical structure factor. A clean data collapse ensures a precise estimation of the independent critical exponents that eventually determine the universality class of the process.
We examine a class of self-organized interface models in random media driven by \emph{extremal dynamics} as discussed in Ref.~\cite{Maslov_1995}. Interestingly, the isotropic version of the model displays anomalous features (a faceted pattern) and belongs to the same universality class as that of the Sneppen interface (model A).
We also introduce and analyze an anisotropic variant of the Sneppen (model A) and suggest the model belongs to the same universality class as the \emph{tensionless} KPZ process~\cite{Cuerno_2022}.

Consider a one-dimensional lattice with site labels $1, 2, \dots L$, along with periodic boundary conditions. To each site, assign the interface height $h(x, t)$. The model-specific update rules are as follows. 

Sneppen~\cite{Sneppen_1992} introduced a discrete model of the kinetic roughening interface in the presence of quenched disorder. The model is a striking example of a self-organized rough interface, showing scale invariance in the height profile. Initially, the interface is flat $h(x, t=0) = 0$. To each site, assign a random pinning force $\eta(x, h(x))$, drawn from a uniform distribution between 0 and 1. The update rules include the following steps: Choose a site $x'$ with the smallest pinning force only among the sites that satisfy the following slope constraints (Kim-Kosterlitz conditions): $[|h(x) +1 - h(x-1)|\leq 1$ and $|h(x) +1 - h(x+1)|\leq 1]$. Then, increase the height of that site by one unit: $h(x', t+1) = h(x', t) +1$.

The model produces a rough interface with faceted pattern. The global width characteristic exponents are $z = 1$ and $\chi = 1$, implying that the process shows \emph{self-similarity}~\cite{Sneppen_1992}. Although the global interface width characteristic exponents seem trivial, the analysis of the dynamical structure factor revealed an unexpectedly nontrivial feature with $\chi_s>\chi$~\cite{Lopez_2000}.

We also examine an anisotropic variant of the model. Here, we implement a different local constraint, $h(x+1)-h(x)\geq 0$. The rest of the dynamical update occurs similarly, as mentioned for the isotropic version of the Sneppen (model A). Eventually, the local slope can have $h(x+1)-h(x) = -1$ or $\gg 1$. While the process keeps the two exponents $z = 1$ and $\chi = 1$ unchanged, the spectral roughness exponent becomes $\chi_s<\chi$ (intrinsically anomalous). Surprisingly, our analysis suggests that the model belongs to the universality class of the tensionless one-dimensional KPZ equation.

Maslov and Zhang~\cite{Maslov_1995} introduced and solved a model of self-organized criticality with a preferred direction. The model is an anisotropic variant of the Zaitsev model~\cite{Zaitsev_1992}. They also suggested physically relevant interface dynamics in random media (quenched disorder) belonging to the same universality class. Our interest is in a variant of the roughening interface model.

In the isotropic version of the model, $F(x) = A[h(x+1) - 2h(x) +h(x-1)] +\eta(x, h(x))$ determines the local force. Here $A$ is the relative strength of the elastic force, and $\eta(x, h(x))$ is a random pinning strength drawn from a uniform distribution between 0 and 1. We use $A=1$ in simulations.
Initially, the interface is a groove: $h(x=2i-1, t=0) = 1$ and $h(x=2i, t=0) = 0$, where $i = 1, 2, \dots, L/2$. At each site, the slope is $h(x+1, t)-h(x, t) = \pm 1$. The local minimum occurs at the value $n_c(x) = h(x+1, t) -2h(x, t)+h(x-1, t) = 2$. In general, $n_c$ can be -2, 0, or 2. The update occurs as $h(x', t+1) = h(x', t) + 2$, where the site $x'$ corresponds to the sites with $n_c = 2$ and having the largest value of the quenched disorder $\eta(x, h)$, or simply the maximum force location for $n_c(x)+\eta(x, h)$.
We call it Maslov-Zhang model B-1, or (MZB-1). As numerically shown below, the model belongs to the same universality class as the Sneppen (model A).

In the anisotropic version of the model (say, MZB-2), the local driving force acting on a site $x$ is $F(x) = A[h(x+1) - h(x)] +\eta(x, h(x))$. We again use $A = 1$ and the same initial condition as mentioned for the isotropic version of the model. Only two height gradients, $h(x+1)-h(x)= \pm 1$, are possible. Updates occur similarly at a site where the force has maximum strength. The suggested exponents are $z = 1$ and $\chi = 1/2$~\cite{Maslov_1995}.
As shown below, the model does not exhibit anomalous features. However, the dynamical exponent takes a slightly different value.

{\it FTS for structure factor $S(k, t)$:}
In terms of the Fourier transform of the height function, $\hat{h}(k, t) = L^{-1/2}\sum_x[h(x, t) - \bar{h}(t)] \exp(ikx)$, one can write an expression for the dynamical structure factor (or power spectrum) $S(k, t) = \langle \hat{h}(k, t) \hat{h}(-k, t)\rangle$. This also reveals the height-height correlation: $G(l, t) = 4/L \sum_{2\pi \le k \le \pi/a_0}  [1-\cos(kl)]S(k, t) \propto \int_{2\pi/L}^{\pi/a_0} (dk/2\pi) [1-\cos(kl)]S(k, t)$, where $a_0$ is the lattice spacing~\cite{Lopez_2000}. If we fix the time $t$, $\bar{h}(t)$ becomes constant, implying that the structure factor of $h$ or $h-\bar{h}$ remains the same.

For a fixed time $t$, the structure factor $S(k, t)$ as a function of wave number $k$  shows typically two distinct regimes. Below a cutoff $k\ll k_0 \sim L^{-1}\sim t^{-1/z}$, the power remains independent of $k$ but increases with time as $\sim t^a$. In the nontrivial wave number regime $k\gg k_0$, the structure factor, in general, can show scaling in both arguments  $\sim 1/k^{\alpha}$ and $\sim t^b$. Now, we can write an expression for the structure factor as a function of the two arguments (wave number and time)
\begin{equation}
S(k, t) \sim \begin{cases} t^a, ~~~~~~~~~~ k\ll t^{-1/z},\\ t^b/k^{\alpha}, ~~~~~~t^{-1/z}\ll k \ll 1/2. \end{cases}
\label{eq_ft_sf1}
\end{equation}
In the regime $k\gg k_0$, the time-dependent scaling of the structure factor along with the scaling function [cf. Eq.~(\ref{eq_ft_sf1})] is
\begin{eqnarray}
S(k, t) \sim \frac{t^b}{k^{\alpha}} \sim  \frac{t^a}{t^{(a-b)}k^{\alpha}} \sim  \frac{t^a}{(kt^{1/z})^{(a-b)z}} \frac{k^{(a-b)z}}{k^{\alpha}} \nonumber \\  \sim  \frac{t^a}{u^{(a-b)z}} \sim t^a H(u),\nonumber
\end{eqnarray}
if the exponent $\alpha$ satisfies a scaling relation
\begin{equation}
 \alpha = (a-b)z.
\label{eq_ft_sf5}
\end{equation}
Similarly, the $k$-dependent scaling of the structure factor along with the scaling function is 
\begin{eqnarray}
S(k, t) \sim t^a H(u) \sim \frac{(kt^{1/z})^{az}H(u)}{k^{az}} \sim G(u)/k^{az},\nonumber
\end{eqnarray}
with $G(u) = u^{az}H(u)$.

With the above scaling arguments, one can express the scaling behavior of the structure factor in one variable with a scaling function in terms of the reduced wave number $u = kt^{1/z}$ as
\begin{equation}
S(k,t) = \frac{1}{k^{az}}G(u) = t^aH(u).
\label{eq_ft_sf2}
\end{equation}
The scaling functions $G(u)$ and $H(u)$ in Eq.~(\ref{eq_ft_sf2}) vary as
\begin{subequations}
\begin{align}
G(u) \sim \begin{cases} u^{az}, ~~~~~~~~~~~~~~u\ll 1,\\ u^{bz}, ~~~~~~~~~~~~~~u\gg 1,\end{cases}
\label{eq_ft_sf3}
\end{align}
\begin{align}
H(u) \sim \begin{cases} {\rm constant}, ~~~~~~~~~~~u\ll 1,\\ 1/u^{(a-b)z}, ~~~~~~~~~~u\gg 1.\end{cases}
\label{eq_ft_sf4}
\end{align}
\end{subequations}

Moreover, the square of the global width,
$w^2(t) \sim \int dk S(k,t) \sim t^{a} \int dk H(u) \sim t^{a-1/z}\sim t^c,$
suggests a scaling relation for the dynamical exponent
 \begin{equation}
1/z = a-c.
\label{eq_ft_sf6}
\end{equation}

Numerically, it is easy to determine the scaling functions: $G(u=kt^{1/z}) = k^{az}S(k,t)$ and $H(u) = t^{-a}S(k,t)$. We only require the two critical exponents $a$ and $z = 1/(a-c)$. One can easily estimate the two exponents by examining the scaling behavior of the low-wave number component power and the square of the global interface width as a function of time.

\begin{figure*}[t]
  \centering
   \scalebox{0.6}{\includegraphics{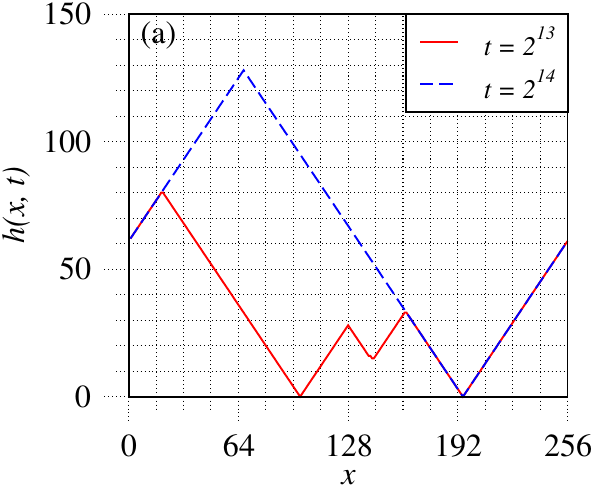}}
  \scalebox{0.6}{\includegraphics{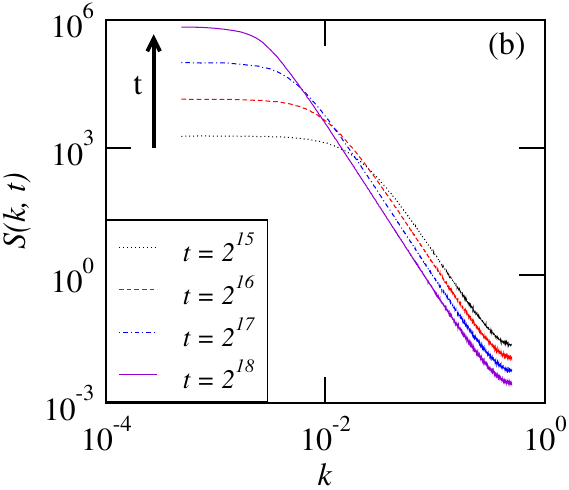}}
   \scalebox{0.6}{\includegraphics{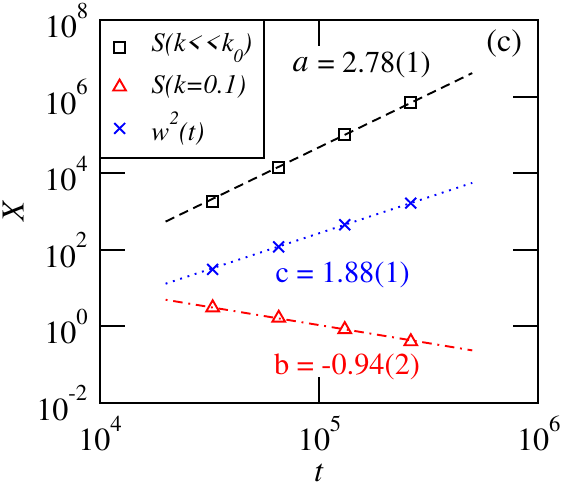}}
     \scalebox{0.6}{\includegraphics{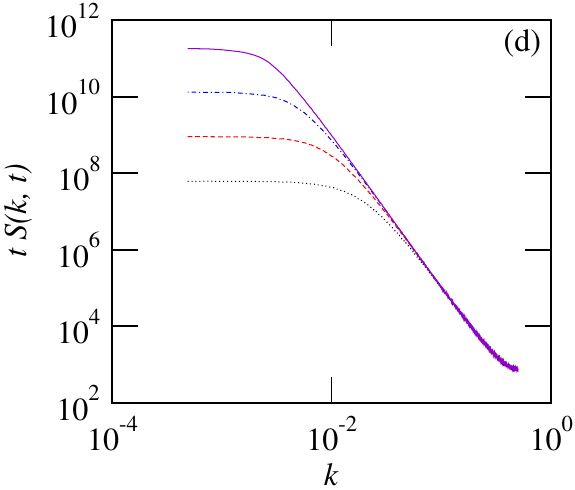}}    
      \scalebox{0.6}{\includegraphics{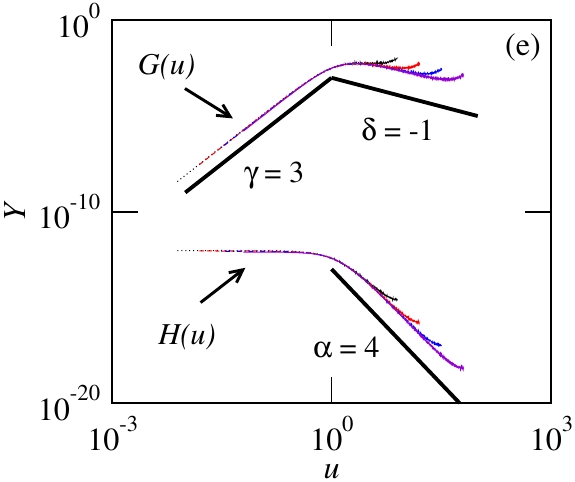}}    
       \scalebox{0.6}{\includegraphics{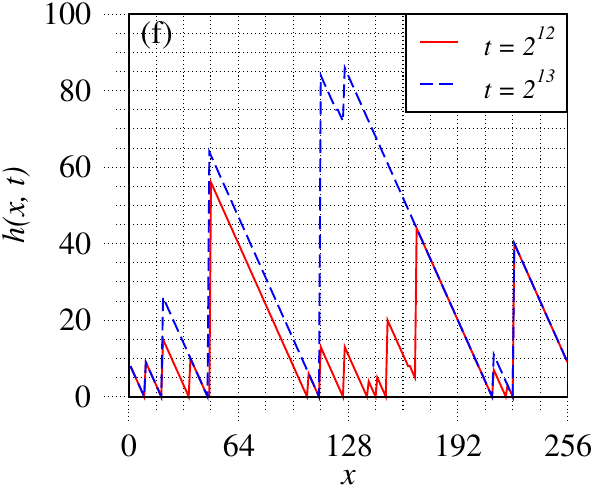}}
       \scalebox{0.6}{\includegraphics{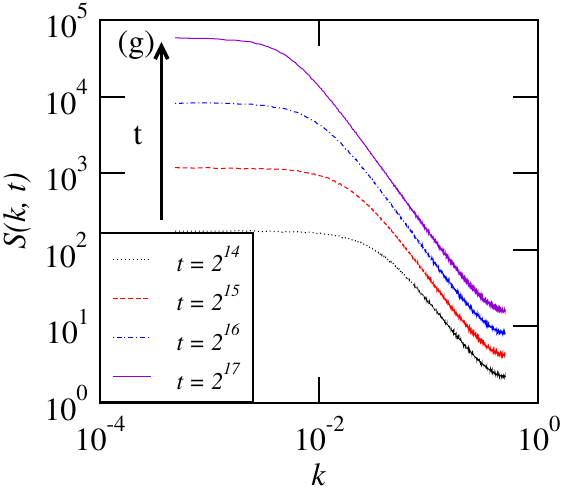}}
      \scalebox{0.6}{\includegraphics{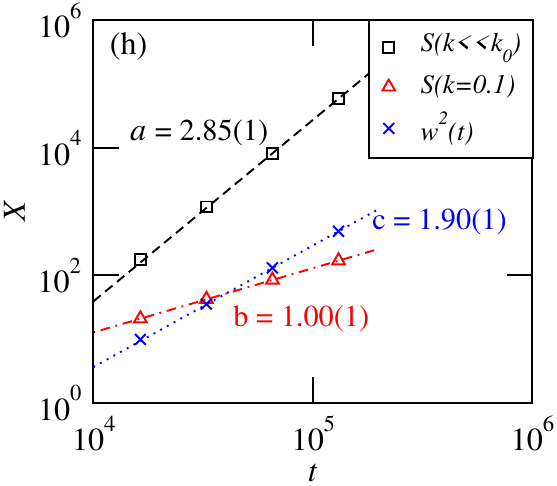}}
      \scalebox{0.6}{\includegraphics{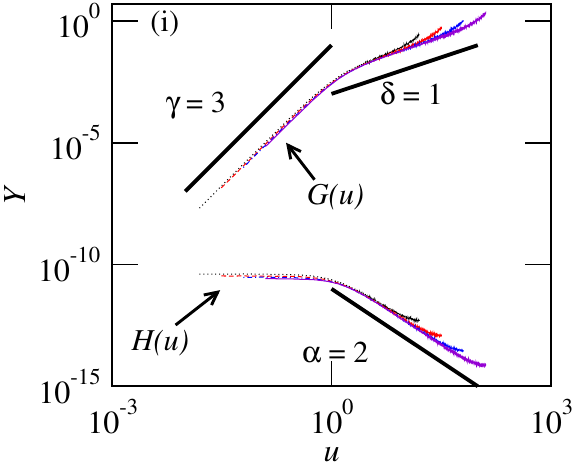}}
  \caption{The Sneppen (model A): (a) Typical height profiles at two values of time. (b) The structure factor $S(k, t)$ after an evolution time $t$ for the rough profile $h(x, t)$, with $L = 2^{12}$. (c) At a fixed wave number $k$, the time scaling of power ($\square$) $\sim t^a$ for $k \ll k_0$ and the power ({\color{red}$\triangle$}) $\sim t^b$ for $k \gg k_0$. The symbol ({\color{blue}$\times$}) corresponds to the square of the global width  $w^2(t) \sim t^c$. Straight lines are the best-fit curves. (d) A clean plot of $t S(k,t)$ vs $k$ shows the absence of time dependence in the nontrivial $k$ regime, or $S(k, t) \sim 1/t$. (e) The scaling functions for the structure factor: $G(u) \sim k^{az}S(k, t)$ and $H(u) \sim t^{-a}S(k, t)$, with an argument $u = kt^{1/z}$. We get finer data collapse using theoretically expected values for the two critical exponents $a = 3$ and $c = 2$. The thick, straight lines guide the slope. Here, the typical characteristic exponents are $z = 1$, $\chi = 1$, and $\chi_s = 3/2$. Similarly, (f)--(i) correspond to an anisotropic variant of the Sneppen (model A), resulting $z = 1$, $\chi = 1$, and $\chi_s = 1/2$.}
  \label{fig_sf_sma}
\end{figure*}

\begin{table*}[t]
\centering
\begin{tabular}{|c|ccc|c|ccc|ccc|}
\hline 
\hline
 ~~~~Model~~~~~~&  ~~ $a$~~   &    ~~$b$~~  & ~~ $c$ ~~~ & ~~~~~~$\beta~~~~~~ $ &~~~~ $z$~~~~ & ~~~$\chi$ ~~~& ~~~~ $\chi_s $~~~~~~& ~~~$\gamma$~~~ & ~~~$\delta$~~~& ~~~$\alpha$~~~~ \\
      &   &  & & $c/2$ & $(a-c)^{-1}$ & $cz/2$ & $(c-b)z/2$ & $az$ & $bz$ & $(a-b)z$  \\
     \hline

    &   3 & -1 & 2 &  1 & 1  & 1 & 3/2 & 3 & -1 & 4 \\
     
Sneppen (model A)&  2.78(1) & -0.94(2)  & 1.88(1) & 0.940(5) &  1.11(2) & 1.04(3)  & 1.57(5) & 3.09(8) & -1.04(5) & 4.1(2)\\
 -do-~\cite{Lopez_2000}  &    &  &  &  1 & 1  & 1 & 1.35 & 3 & -0.7 & 3.7 \\
      MZB-1 &  2.78(1) & -0.94(2)  & 1.89(1) & 0.945(5) &  1.12(3) & 1.06(3)  & 1.59(5) & 3.12(8) & -1.05(5) & 4.2(2) \\
    
\hline
    Anisotropic &   3 & 1 & 2 &  1 & 1  & 1 & 1/2 & 3 & 1 & 2 \\
     Sneppen (model A) &  2.85(1) & 1.00(1)  & 1.90(1) & 0.950(5) &  1.05(2) & 1.00(3) & 0.47(2) & 3.00(7) & 1.05(3) & 1.95(5) \\
    \hline
      MZB-2 &  3.22(1) & 0  & 1.69(1) & 0.845(5) &  0.65(1) & 0.55(1)  & 0.55(1) & 2.10(3) & 0 & 2.10(3) \\

\hline
\hline
\end{tabular}
 \caption{A summary of the critical exponents characterizing the dynamical structure factor properties. For Sneppen (model A), the first row presents expected theoretical exponents.}
\label{tab1}
\end{table*}

For comparison, we write the scaling ansatz for the structure factor (in spatial dimension $d = 1$) as proposed in the previous studies,
\begin{equation}
S(k, t) =k^{-(2\chi+1)}G(kt^{1/z}) = t^{(2\chi+1)/z}H(u),
\label{eq_sf01}
\end{equation}
where the most general form of the scaling functions in Eq.~(\ref{eq_sf01}) are
\begin{subequations}
\begin{align}
G(u) \sim \begin{cases} u^{2\chi+1}, ~~~~~~{\rm if}~~~u\ll 1,\\ u^{2(\chi-\chi_s)}, ~~~{\rm if}~~~u\gg 1, \end{cases}
\label{eq_sf02}
\end{align}
\begin{align}
H(u) \sim \begin{cases} {\rm constant}, ~~~~{\rm if}~~~u\ll 1,\\ u^{-(2\chi_s+1)}, ~~~{\rm if}~~~u\gg 1. \end{cases}
\label{eq_sf03}
\end{align}
\end{subequations}
The exponent $\chi_s$ is the spectral roughness exponent. It is easy to note that no trace of the spectral roughness exponent appears in the global width, as $w^2(t) = \int dk S(k, t) = t^{(2\chi+1)/z}\int H(u) d(k t^{1/z})/t^{1/z} \sim t^{2\chi/z} \sim t^{2\beta}$. Because of this the structure factor is the most relevant characterization and provides subtle details of the underlying process.

It is easy to recognize the following scaling relations in terms of the exponents $a, b$, and/or $c$. The global interface width characteristic exponents are the growth exponent $\beta = c/2$, the dynamical exponent $z = 1/(a-c)$ [cf. Eq.~(\ref{eq_ft_sf6})], and the global roughness exponent $\chi = \beta z  = cz/2$. For the scaling function $G(u)$ [cf. Eqs.~(\ref{eq_ft_sf3}) and ~(\ref{eq_sf02})], the critical exponents are 
\begin{eqnarray}
\gamma = 2\chi + 1 = az ~~~{\rm and}~~~\delta = 2(\chi-\chi_s) = bz.\nonumber
\end{eqnarray}
Similarly, the critical exponent for the scaling function $H(u)$ [cf. Eqs.~(\ref{eq_ft_sf5}), (\ref{eq_ft_sf4}), and ~(\ref{eq_sf03})] is 
\begin{equation}
\alpha = \gamma-\delta = 2\chi_s+1 = (a-b)z. \nonumber
\end{equation}
This is related to the spectral roughness exponent 
\begin{equation}
\chi_s = \chi-bz/2 = (c-b)z/2. 
\label{eq_chis3}
\end{equation}
 Eq.~(\ref{eq_chis3}) implies that
 \begin{eqnarray}
 b = 0 \Rightarrow \chi_s = \chi ~~~~ \begin{cases} {\rm if}~\chi_s < 1  ~~  {\Rightarrow\rm ~~ Family~ Vicsek}\\  {\rm if}~\chi_s > 1 ~~ {\Rightarrow  \rm ~~ super~rough},  \end{cases} \nonumber\\
 b>0 \Rightarrow \chi_s < \chi  ~~ \Rightarrow ~~ {\rm intrinsic~anomalous}, \nonumber\\ 
 b<0 \Rightarrow \chi_s > \chi  ~~ \Rightarrow ~~ {\rm faceted~pattern.~~~~~~} \nonumber 
 \end{eqnarray}

Figure~\ref{fig_sf_sma} displays the properties of the dynamical structure factor and its analysis using FTS for the Sneppen (model A).  Table~\ref{tab1} presents the estimated critical exponents. Similarly, we studied the MZB-1 and MZB-2 models. Clean data collapse excellently supports the numerically estimated exponents within the statistical error.
Our results are consistent with Refs.~\cite{Sneppen_1992, Lopez_2000} that suggest $z = 1$ and $\chi = 1$  for the Sneppen (model A). This implies that $a = 3$ and $c = 2$, as $2\chi + 1 = az$ and $z = 1/(a-c)$. Further, our finer numerical results [cf. Fig.~\ref{fig_sf_sma}(d)] suggest $b = -1$. Eventually, the spectral roughness exponent is $\chi_s = (c-b)z/2 = 3/2$, which differs slightly from the previously estimated value of 1.35~\cite{Lopez_2000}. We also get the same set of exponents for the MZB-1 model, indicating that the two models belong to the same universality class.

Similarly, the critical exponents for the anisotropic variant of the Sneppen (model A) are $z = 1$, $\chi = 1$, and $\chi_s  = 1/2$ [cf. Fig.~\ref{fig_sf_sma}(i)], implying the model shows intrinsically anomalous behavior.
More recently, Rodr\'iguez-Fern\'andez \etl \cite{Cuerno_2022} provided a direct numerical simulation of the tensionless one-dimensional KPZ equation that shows intrinsic anomalous behavior with $z = 1$, $\chi = 1$, and $\chi_s = 1/2$.  One can describe the space derivative of the height profile $u = \partial_x h$ by the inviscid stochastic Burger equation $\partial_t u = \lambda_2 u \partial_x u + \partial_x \eta(x, t)$, where $\eta(x, t)$ denotes uncorrelated noise in space-time. Assuming $u$ as a rough interface, they examined its dynamical structure factor and found $z = 2/3$, $\chi = 1/3$, and $\chi_s = -1/2$. They also examined the stochastic Korteweg-de Vries (KdV) equation (an important model of weakly nonlinear waves)  $\partial_t u = c \partial_{x}^{3}u + u \partial_x u + \partial_x \eta(x, t)$, which one can get from  $\partial_t h = c \partial_{x}^{3}h + (\partial_x h)^2/2 + \eta(x, t)$ with $u = \partial_x h$. Interestingly, they observed that the height profile corresponding to the stochastic KdV equation also belongs to the same universality class as that of the tensionless one-dimensional KPZ equation. Similarly, the stochastic KdV equation and the inviscid stochastic Burger equation belong to the same universality. Our results suggest that the anisotropic version of the Sneppen (model A) and the tensionless one-dimensional KPZ equation seem to share the same universality class.

As seen from Fig.~\ref{fig_sf_sma}(f), the height profile for the anisotropic case does not satisfy the space inversion symmetry ($x\to -x$). Breaking the spatial inversion symmetry [going from an isotropic to an anisotropic variant of the Sneppen (model A)] can change the universality class, implying a transition from a subclass with a faceted pattern to a subclass with an intrinsic anomalous feature for the kinetic roughening interfaces.  
For the MZB-2 model, the expected exponents, as mentioned in Ref.~\cite{Maslov_1995} are $z = 1$ and $\chi = 1/2$. However, our numerical result (not shown) suggests ($\chi_s = \chi < 1$), with $z \approx 2/3$ and $\chi \approx 1/2$.


In summary, we have provided a \emph{systematic} finite-time scaling for the dynamical structure factor. We emphasized that the method can accurately determine the \emph{universality features} (the critical exponents and the scaling functions). In fact, the approach is general and applicable to a wide range of rough surfaces or interfaces. As an illustration, we have applied the method to a class of discrete models [Sneppen (model A) and Maslov-Zhang model B] of rough interfaces in the presence of quenched disorder driven by extremal dynamics. 

Finally, we highlight the interesting conclusion of our simulation studies.
(i) The MZB-1 model shows faceted patterns ($\chi_s > \chi$), with $z = 1$, $\chi = 1$, and $\chi_s = 3/2$. Strikingly, the model belongs to the same universality class as that of the Sneppen (model A). Experimentally, such behavior has been found in the kinetic roughening of dissolving polycrystalline pure iron~\cite{Bastos_2009}. (ii) We also introduced and examined an anisotropic variant of the Sneppen (model A). The model shows an intrinsically anomalous feature ($\chi_s < \chi$), with $z = 1$, $\chi = 1$, and $\chi_s = 1/2$, and seems to belong to the universality class of the tensionless one-dimensional KPZ equation or the height profile corresponding to the stochastic KdV equation~\cite{Cuerno_2022}.
(iii) In particular, the MZB-2 model has been of interest in the context of a solvable model of self-organized criticality (SOC)~\cite{Maslov_1995}. Although the MZB-2 model exhibits Family-Vicsek type scaling ($\chi = \chi_s < 1$) with $z \approx 2/3$ and $\chi \approx 1/2$, the dynamical exponent significantly differs from the previously argued value $z = 1$~\cite{Maslov_1995}. Therefore, the MZB-2 model does not belong to the same universality class as the SOC model discussed in Ref.~\cite{Maslov_1995}.
We have also examined several discrete models of standard universality and consistently found the applicability of the FTS method.
Although the FTS can enhance our understanding significantly, the entire set of physical features that determine anomalous behavior needs further exploration.

\textit{Acknowledgments}. RC acknowledges financial support through the Junior Research Fellowship, UGC, India. ACY acknowledges a seed grant under IoE by Banaras Hindu University (Seed Grant-II/2022-23/48729).

\balance
\bibliography{papersources}

\begin{thebibliography}{44}%
\makeatletter
\providecommand \@ifxundefined [1]{%
 \@ifx{#1\undefined}
}%
\providecommand \@ifnum [1]{%
 \ifnum #1\expandafter \@firstoftwo
 \else \expandafter \@secondoftwo
 \fi
}%
\providecommand \@ifx [1]{%
 \ifx #1\expandafter \@firstoftwo
 \else \expandafter \@secondoftwo
 \fi
}%
\providecommand \natexlab [1]{#1}%
\providecommand \enquote  [1]{``#1''}%
\providecommand \bibnamefont  [1]{#1}%
\providecommand \bibfnamefont [1]{#1}%
\providecommand \citenamefont [1]{#1}%
\providecommand \href@noop [0]{\@secondoftwo}%
\providecommand \href [0]{\begingroup \@sanitize@url \@href}%
\providecommand \@href[1]{\@@startlink{#1}\@@href}%
\providecommand \@@href[1]{\endgroup#1\@@endlink}%
\providecommand \@sanitize@url [0]{\catcode `\\12\catcode `\$12\catcode
  `\&12\catcode `\#12\catcode `\^12\catcode `\_12\catcode `\%12\relax}%
\providecommand \@@startlink[1]{}%
\providecommand \@@endlink[0]{}%
\providecommand \url  [0]{\begingroup\@sanitize@url \@url }%
\providecommand \@url [1]{\endgroup\@href {#1}{\urlprefix }}%
\providecommand \urlprefix  [0]{URL }%
\providecommand \Eprint [0]{\href }%
\providecommand \doibase [0]{http://dx.doi.org/}%
\providecommand \selectlanguage [0]{\@gobble}%
\providecommand \bibinfo  [0]{\@secondoftwo}%
\providecommand \bibfield  [0]{\@secondoftwo}%
\providecommand \translation [1]{[#1]}%
\providecommand \BibitemOpen [0]{}%
\providecommand \bibitemStop [0]{}%
\providecommand \bibitemNoStop [0]{.\EOS\space}%
\providecommand \EOS [0]{\spacefactor3000\relax}%
\providecommand \BibitemShut  [1]{\csname bibitem#1\endcsname}%
\let\auto@bib@innerbib\@empty
\bibitem [{\citenamefont {Halpin-Healy}\ and\ \citenamefont
  {Zhang}(1995)}]{Zhang_1995}%
  \BibitemOpen
  \bibfield  {author} {\bibinfo {author} {\bibfnamefont {T.}~\bibnamefont
  {Halpin-Healy}}\ and\ \bibinfo {author} {\bibfnamefont {Y.-C.}\ \bibnamefont
  {Zhang}},\ }\href {\doibase https://doi.org/10.1016/0370-1573(94)00087-J}
  {\bibfield  {journal} {\bibinfo  {journal} {Phys. Rep.}\ }\textbf {\bibinfo
  {volume} {254}},\ \bibinfo {pages} {215} (\bibinfo {year}
  {1995})}\BibitemShut {NoStop}%
\bibitem [{\citenamefont {Barab\'asi}\ and\ \citenamefont
  {Stanley}(1995)}]{Stanley_1995}%
  \BibitemOpen
  \bibfield  {author} {\bibinfo {author} {\bibfnamefont {A.-L.}\ \bibnamefont
  {Barab\'asi}}\ and\ \bibinfo {author} {\bibfnamefont {H.~E.}\ \bibnamefont
  {Stanley}},\ }\href@noop {} {\emph {\bibinfo {title} {Fractal Concepts in
  Surface Growth}}}\ (\bibinfo  {publisher} {Cambridge University Press,
  Cambridge, England},\ \bibinfo {year} {1995})\BibitemShut {NoStop}%
\bibitem [{\citenamefont {T\"auber}(2014)}]{Tauber_2014}%
  \BibitemOpen
  \bibfield  {author} {\bibinfo {author} {\bibfnamefont {U.~C.}\ \bibnamefont
  {T\"auber}},\ }\href@noop {} {\emph {\bibinfo {title} {Critical Dynamics: A
  Field Theory Approach to Equilibrium and Non-Equilibrium Scaling Behavior}}}\
  (\bibinfo  {publisher} {Cambridge University Press, Cambridge, U.K.},\
  \bibinfo {year} {2014})\BibitemShut {NoStop}%
\bibitem [{\citenamefont {Forgerini}\ and\ \citenamefont
  {Marchiori}(2014)}]{Marchiori_2014}%
  \BibitemOpen
  \bibfield  {author} {\bibinfo {author} {\bibfnamefont {F.~L.}\ \bibnamefont
  {Forgerini}}\ and\ \bibinfo {author} {\bibfnamefont {R.}~\bibnamefont
  {Marchiori}},\ }\href {\doibase 10.4161/biom.28871} {\bibfield  {journal}
  {\bibinfo  {journal} {Biomatter}\ }\textbf {\bibinfo {volume} {4}},\ \bibinfo
  {pages} {e28871} (\bibinfo {year} {2014})}\BibitemShut {NoStop}%
\bibitem [{\citenamefont {Soriano}\ \emph {et~al.}(2002)\citenamefont
  {Soriano}, \citenamefont {Ramasco}, \citenamefont {Rodr\'{\i}guez},
  \citenamefont {Hern\'andez-Machado},\ and\ \citenamefont
  {Ort\'{\i}n}}]{Soriano_2002}%
  \BibitemOpen
  \bibfield  {author} {\bibinfo {author} {\bibfnamefont {J.}~\bibnamefont
  {Soriano}}, \bibinfo {author} {\bibfnamefont {J.~J.}\ \bibnamefont
  {Ramasco}}, \bibinfo {author} {\bibfnamefont {M.~A.}\ \bibnamefont
  {Rodr\'{\i}guez}}, \bibinfo {author} {\bibfnamefont {A.}~\bibnamefont
  {Hern\'andez-Machado}}, \ and\ \bibinfo {author} {\bibfnamefont
  {J.}~\bibnamefont {Ort\'{\i}n}},\ }\href
  {https://link.aps.org/doi/10.1103/PhysRevLett.89.026102} {\bibfield
  {journal} {\bibinfo  {journal} {Phys. Rev. Lett.}\ }\textbf {\bibinfo
  {volume} {89}},\ \bibinfo {pages} {026102} (\bibinfo {year}
  {2002})}\BibitemShut {NoStop}%
\bibitem [{\citenamefont {L\'opez}\ and\ \citenamefont
  {Schmittbuhl}(1998)}]{Schmittbuhl_1998}%
  \BibitemOpen
  \bibfield  {author} {\bibinfo {author} {\bibfnamefont {J.~M.}\ \bibnamefont
  {L\'opez}}\ and\ \bibinfo {author} {\bibfnamefont {J.}~\bibnamefont
  {Schmittbuhl}},\ }\href {https://link.aps.org/doi/10.1103/PhysRevE.57.6405}
  {\bibfield  {journal} {\bibinfo  {journal} {Phys. Rev. E}\ }\textbf {\bibinfo
  {volume} {57}},\ \bibinfo {pages} {6405} (\bibinfo {year}
  {1998})}\BibitemShut {NoStop}%
\bibitem [{\citenamefont {Ponson}\ \emph {et~al.}(2006)\citenamefont {Ponson},
  \citenamefont {Bonamy},\ and\ \citenamefont {Bouchaud}}]{Ponson_2006}%
  \BibitemOpen
  \bibfield  {author} {\bibinfo {author} {\bibfnamefont {L.}~\bibnamefont
  {Ponson}}, \bibinfo {author} {\bibfnamefont {D.}~\bibnamefont {Bonamy}}, \
  and\ \bibinfo {author} {\bibfnamefont {E.}~\bibnamefont {Bouchaud}},\ }\href
  {https://link.aps.org/doi/10.1103/PhysRevLett.96.035506} {\bibfield
  {journal} {\bibinfo  {journal} {Phys. Rev. Lett.}\ }\textbf {\bibinfo
  {volume} {96}},\ \bibinfo {pages} {035506} (\bibinfo {year}
  {2006})}\BibitemShut {NoStop}%
\bibitem [{\citenamefont {L\'opez}\ and\ \citenamefont
  {Jensen}(1998)}]{Jensen_1998}%
  \BibitemOpen
  \bibfield  {author} {\bibinfo {author} {\bibfnamefont {J.~M.}\ \bibnamefont
  {L\'opez}}\ and\ \bibinfo {author} {\bibfnamefont {H.~J.}\ \bibnamefont
  {Jensen}},\ }\href {https://link.aps.org/doi/10.1103/PhysRevLett.81.1734}
  {\bibfield  {journal} {\bibinfo  {journal} {Phys. Rev. Lett.}\ }\textbf
  {\bibinfo {volume} {81}},\ \bibinfo {pages} {1734} (\bibinfo {year}
  {1998})}\BibitemShut {NoStop}%
\bibitem [{\citenamefont {Das~Sarma}\ and\ \citenamefont
  {Tamborenea}(1991)}]{Sarma_1991}%
  \BibitemOpen
  \bibfield  {author} {\bibinfo {author} {\bibfnamefont {S.}~\bibnamefont
  {Das~Sarma}}\ and\ \bibinfo {author} {\bibfnamefont {P.}~\bibnamefont
  {Tamborenea}},\ }\href {https://link.aps.org/doi/10.1103/PhysRevLett.66.325}
  {\bibfield  {journal} {\bibinfo  {journal} {Phys. Rev. Lett.}\ }\textbf
  {\bibinfo {volume} {66}},\ \bibinfo {pages} {325} (\bibinfo {year}
  {1991})}\BibitemShut {NoStop}%
\bibitem [{\citenamefont {Lai}\ and\ \citenamefont
  {Das~Sarma}(1991)}]{Lai_1991}%
  \BibitemOpen
  \bibfield  {author} {\bibinfo {author} {\bibfnamefont {Z.-W.}\ \bibnamefont
  {Lai}}\ and\ \bibinfo {author} {\bibfnamefont {S.}~\bibnamefont
  {Das~Sarma}},\ }\href {https://link.aps.org/doi/10.1103/PhysRevLett.66.2348}
  {\bibfield  {journal} {\bibinfo  {journal} {Phys. Rev. Lett.}\ }\textbf
  {\bibinfo {volume} {66}},\ \bibinfo {pages} {2348} (\bibinfo {year}
  {1991})}\BibitemShut {NoStop}%
\bibitem [{\citenamefont {Kim}\ and\ \citenamefont
  {Das~Sarma}(1994)}]{Kim_1994}%
  \BibitemOpen
  \bibfield  {author} {\bibinfo {author} {\bibfnamefont {J.~M.}\ \bibnamefont
  {Kim}}\ and\ \bibinfo {author} {\bibfnamefont {S.}~\bibnamefont
  {Das~Sarma}},\ }\href {https://link.aps.org/doi/10.1103/PhysRevLett.72.2903}
  {\bibfield  {journal} {\bibinfo  {journal} {Phys. Rev. Lett.}\ }\textbf
  {\bibinfo {volume} {72}},\ \bibinfo {pages} {2903} (\bibinfo {year}
  {1994})}\BibitemShut {NoStop}%
\bibitem [{\citenamefont {Edwards}\ and\ \citenamefont
  {Wilkinson}(1982)}]{Edwards_1982}%
  \BibitemOpen
  \bibfield  {author} {\bibinfo {author} {\bibfnamefont {S.~F.}\ \bibnamefont
  {Edwards}}\ and\ \bibinfo {author} {\bibfnamefont {D.~R.}\ \bibnamefont
  {Wilkinson}},\ }\href {\doibase 10.1098/rspa.1982.0056} {\bibfield  {journal}
  {\bibinfo  {journal} {Proc. R. Soc. Lond. A}\ }\textbf {\bibinfo {volume}
  {381}},\ \bibinfo {pages} {17} (\bibinfo {year} {1982})}\BibitemShut
  {NoStop}%
\bibitem [{\citenamefont {Kardar}\ \emph {et~al.}(1986)\citenamefont {Kardar},
  \citenamefont {Parisi},\ and\ \citenamefont {Zhang}}]{Kardar_1986}%
  \BibitemOpen
  \bibfield  {author} {\bibinfo {author} {\bibfnamefont {M.}~\bibnamefont
  {Kardar}}, \bibinfo {author} {\bibfnamefont {G.}~\bibnamefont {Parisi}}, \
  and\ \bibinfo {author} {\bibfnamefont {Y.-C.}\ \bibnamefont {Zhang}},\ }\href
  {https://link.aps.org/doi/10.1103/PhysRevLett.56.889} {\bibfield  {journal}
  {\bibinfo  {journal} {Phys. Rev. Lett.}\ }\textbf {\bibinfo {volume} {56}},\
  \bibinfo {pages} {889} (\bibinfo {year} {1986})}\BibitemShut {NoStop}%
\bibitem [{\citenamefont {Sasamoto}\ and\ \citenamefont
  {Spohn}(2010)}]{Sasamoto_2010}%
  \BibitemOpen
  \bibfield  {author} {\bibinfo {author} {\bibfnamefont {T.}~\bibnamefont
  {Sasamoto}}\ and\ \bibinfo {author} {\bibfnamefont {H.}~\bibnamefont
  {Spohn}},\ }\href {https://link.aps.org/doi/10.1103/PhysRevLett.104.230602}
  {\bibfield  {journal} {\bibinfo  {journal} {Phys. Rev. Lett.}\ }\textbf
  {\bibinfo {volume} {104}},\ \bibinfo {pages} {230602} (\bibinfo {year}
  {2010})}\BibitemShut {NoStop}%
\bibitem [{\citenamefont {Calabrese}\ and\ \citenamefont
  {Le~Doussal}(2011)}]{Calabrese_2011}%
  \BibitemOpen
  \bibfield  {author} {\bibinfo {author} {\bibfnamefont {P.}~\bibnamefont
  {Calabrese}}\ and\ \bibinfo {author} {\bibfnamefont {P.}~\bibnamefont
  {Le~Doussal}},\ }\href
  {https://link.aps.org/doi/10.1103/PhysRevLett.106.250603} {\bibfield
  {journal} {\bibinfo  {journal} {Phys. Rev. Lett.}\ }\textbf {\bibinfo
  {volume} {106}},\ \bibinfo {pages} {250603} (\bibinfo {year}
  {2011})}\BibitemShut {NoStop}%
\bibitem [{\citenamefont {Takeuchi}(2018)}]{Takeuchi_2018}%
  \BibitemOpen
  \bibfield  {author} {\bibinfo {author} {\bibfnamefont {K.~A.}\ \bibnamefont
  {Takeuchi}},\ }\href {\doibase https://doi.org/10.1016/j.physa.2018.03.009}
  {\bibfield  {journal} {\bibinfo  {journal} {Physica A}\ }\textbf {\bibinfo
  {volume} {504}},\ \bibinfo {pages} {77} (\bibinfo {year} {2018})}\BibitemShut
  {NoStop}%
\bibitem [{\citenamefont {Guti\'errez}\ and\ \citenamefont
  {Cuerno}(2023)}]{Cuerno_2023}%
  \BibitemOpen
  \bibfield  {author} {\bibinfo {author} {\bibfnamefont {R.}~\bibnamefont
  {Guti\'errez}}\ and\ \bibinfo {author} {\bibfnamefont {R.}~\bibnamefont
  {Cuerno}},\ }\href
  {https://link.aps.org/doi/10.1103/PhysRevResearch.5.023047} {\bibfield
  {journal} {\bibinfo  {journal} {Phys. Rev. Res.}\ }\textbf {\bibinfo {volume}
  {5}},\ \bibinfo {pages} {023047} (\bibinfo {year} {2023})}\BibitemShut
  {NoStop}%
\bibitem [{\citenamefont {Family}(1986)}]{Family_1986}%
  \BibitemOpen
  \bibfield  {author} {\bibinfo {author} {\bibfnamefont {F.}~\bibnamefont
  {Family}},\ }\href {\doibase 10.1088/0305-4470/19/8/006} {\bibfield
  {journal} {\bibinfo  {journal} {J. Phys. A: Math. Gen.}\ }\textbf {\bibinfo
  {volume} {19}},\ \bibinfo {pages} {L441} (\bibinfo {year}
  {1986})}\BibitemShut {NoStop}%
\bibitem [{\citenamefont {Kwak}\ and\ \citenamefont {Kim}(2019)}]{Kwak_2019}%
  \BibitemOpen
  \bibfield  {author} {\bibinfo {author} {\bibfnamefont {W.}~\bibnamefont
  {Kwak}}\ and\ \bibinfo {author} {\bibfnamefont {J.~M.}\ \bibnamefont {Kim}},\
  }\href {\doibase https://doi.org/10.1016/j.physa.2019.01.016} {\bibfield
  {journal} {\bibinfo  {journal} {Physica A}\ }\textbf {\bibinfo {volume}
  {520}},\ \bibinfo {pages} {87} (\bibinfo {year} {2019})}\BibitemShut
  {NoStop}%
\bibitem [{\citenamefont {Eden}(1958)}]{Eden_1958}%
  \BibitemOpen
  \bibfield  {author} {\bibinfo {author} {\bibfnamefont {M.}~\bibnamefont
  {Eden}},\ }\href@noop {} {\emph {\bibinfo {title} {Symposium on Information
  Theory in Biology}}}\ (\bibinfo  {publisher} {Pergamon Press, New York},\
  \bibinfo {year} {1958})\BibitemShut {NoStop}%
\bibitem [{\citenamefont {Family}\ and\ \citenamefont
  {Vicsek}(1985)}]{Family_1985}%
  \BibitemOpen
  \bibfield  {author} {\bibinfo {author} {\bibfnamefont {F.}~\bibnamefont
  {Family}}\ and\ \bibinfo {author} {\bibfnamefont {T.}~\bibnamefont
  {Vicsek}},\ }\href {\doibase 10.1088/0305-4470/18/2/005} {\bibfield
  {journal} {\bibinfo  {journal} {J. Phys. A: Math. Gen.}\ }\textbf {\bibinfo
  {volume} {18}},\ \bibinfo {pages} {L75} (\bibinfo {year} {1985})}\BibitemShut
  {NoStop}%
\bibitem [{\citenamefont {Kim}\ and\ \citenamefont
  {Kosterlitz}(1989)}]{Kim_1989}%
  \BibitemOpen
  \bibfield  {author} {\bibinfo {author} {\bibfnamefont {J.~M.}\ \bibnamefont
  {Kim}}\ and\ \bibinfo {author} {\bibfnamefont {J.~M.}\ \bibnamefont
  {Kosterlitz}},\ }\href {https://link.aps.org/doi/10.1103/PhysRevLett.62.2289}
  {\bibfield  {journal} {\bibinfo  {journal} {Phys. Rev. Lett.}\ }\textbf
  {\bibinfo {volume} {62}},\ \bibinfo {pages} {2289} (\bibinfo {year}
  {1989})}\BibitemShut {NoStop}%
\bibitem [{\citenamefont {Takeuchi}\ and\ \citenamefont
  {Sano}(2010)}]{Takeuchi_2010}%
  \BibitemOpen
  \bibfield  {author} {\bibinfo {author} {\bibfnamefont {K.~A.}\ \bibnamefont
  {Takeuchi}}\ and\ \bibinfo {author} {\bibfnamefont {M.}~\bibnamefont
  {Sano}},\ }\href {https://link.aps.org/doi/10.1103/PhysRevLett.104.230601}
  {\bibfield  {journal} {\bibinfo  {journal} {Phys. Rev. Lett.}\ }\textbf
  {\bibinfo {volume} {104}},\ \bibinfo {pages} {230601} (\bibinfo {year}
  {2010})}\BibitemShut {NoStop}%
\bibitem [{\citenamefont {Takeuchi}\ \emph {et~al.}(2011)\citenamefont
  {Takeuchi}, \citenamefont {Sano}, \citenamefont {Sasamoto},\ and\
  \citenamefont {Spohn}}]{Takeuchi_2011}%
  \BibitemOpen
  \bibfield  {author} {\bibinfo {author} {\bibfnamefont {K.~A.}\ \bibnamefont
  {Takeuchi}}, \bibinfo {author} {\bibfnamefont {M.}~\bibnamefont {Sano}},
  \bibinfo {author} {\bibfnamefont {T.}~\bibnamefont {Sasamoto}}, \ and\
  \bibinfo {author} {\bibfnamefont {H.}~\bibnamefont {Spohn}},\ }\href
  {https://doi.org/10.1038/srep00034} {\bibfield  {journal} {\bibinfo
  {journal} {Sci. Rep.}\ }\textbf {\bibinfo {volume} {1}},\ \bibinfo {pages}
  {34} (\bibinfo {year} {2011})}\BibitemShut {NoStop}%
\bibitem [{\citenamefont {Foltin}\ \emph {et~al.}(1994)\citenamefont {Foltin},
  \citenamefont {Oerding}, \citenamefont {R\'acz}, \citenamefont {Workman},\
  and\ \citenamefont {Zia}}]{Foltin_1994}%
  \BibitemOpen
  \bibfield  {author} {\bibinfo {author} {\bibfnamefont {G.}~\bibnamefont
  {Foltin}}, \bibinfo {author} {\bibfnamefont {K.}~\bibnamefont {Oerding}},
  \bibinfo {author} {\bibfnamefont {Z.}~\bibnamefont {R\'acz}}, \bibinfo
  {author} {\bibfnamefont {R.~L.}\ \bibnamefont {Workman}}, \ and\ \bibinfo
  {author} {\bibfnamefont {R.~K.~P.}\ \bibnamefont {Zia}},\ }\href
  {https://link.aps.org/doi/10.1103/PhysRevE.50.R639} {\bibfield  {journal}
  {\bibinfo  {journal} {Phys. Rev. E}\ }\textbf {\bibinfo {volume} {50}},\
  \bibinfo {pages} {R639} (\bibinfo {year} {1994})}\BibitemShut {NoStop}%
\bibitem [{\citenamefont {Raychaudhuri}\ \emph {et~al.}(2001)\citenamefont
  {Raychaudhuri}, \citenamefont {Cranston}, \citenamefont {Przybyla},\ and\
  \citenamefont {Shapir}}]{Shapir_2001}%
  \BibitemOpen
  \bibfield  {author} {\bibinfo {author} {\bibfnamefont {S.}~\bibnamefont
  {Raychaudhuri}}, \bibinfo {author} {\bibfnamefont {M.}~\bibnamefont
  {Cranston}}, \bibinfo {author} {\bibfnamefont {C.}~\bibnamefont {Przybyla}},
  \ and\ \bibinfo {author} {\bibfnamefont {Y.}~\bibnamefont {Shapir}},\ }\href
  {https://link.aps.org/doi/10.1103/PhysRevLett.87.136101} {\bibfield
  {journal} {\bibinfo  {journal} {Phys. Rev. Lett.}\ }\textbf {\bibinfo
  {volume} {87}},\ \bibinfo {pages} {136101} (\bibinfo {year}
  {2001})}\BibitemShut {NoStop}%
\bibitem [{\citenamefont {Toroczkai}\ \emph {et~al.}(2000)\citenamefont
  {Toroczkai}, \citenamefont {Korniss}, \citenamefont {Das~Sarma},\ and\
  \citenamefont {Zia}}]{Toroczkai_2000}%
  \BibitemOpen
  \bibfield  {author} {\bibinfo {author} {\bibfnamefont {Z.}~\bibnamefont
  {Toroczkai}}, \bibinfo {author} {\bibfnamefont {G.}~\bibnamefont {Korniss}},
  \bibinfo {author} {\bibfnamefont {S.}~\bibnamefont {Das~Sarma}}, \ and\
  \bibinfo {author} {\bibfnamefont {R.~K.~P.}\ \bibnamefont {Zia}},\ }\href
  {https://link.aps.org/doi/10.1103/PhysRevE.62.276} {\bibfield  {journal}
  {\bibinfo  {journal} {Phys. Rev. E}\ }\textbf {\bibinfo {volume} {62}},\
  \bibinfo {pages} {276} (\bibinfo {year} {2000})}\BibitemShut {NoStop}%
\bibitem [{\citenamefont {Shapir}\ \emph {et~al.}(2000)\citenamefont {Shapir},
  \citenamefont {Raychaudhuri}, \citenamefont {Foster},\ and\ \citenamefont
  {Jorne}}]{Shapir_2000}%
  \BibitemOpen
  \bibfield  {author} {\bibinfo {author} {\bibfnamefont {Y.}~\bibnamefont
  {Shapir}}, \bibinfo {author} {\bibfnamefont {S.}~\bibnamefont
  {Raychaudhuri}}, \bibinfo {author} {\bibfnamefont {D.~G.}\ \bibnamefont
  {Foster}}, \ and\ \bibinfo {author} {\bibfnamefont {J.}~\bibnamefont
  {Jorne}},\ }\href {https://link.aps.org/doi/10.1103/PhysRevLett.84.3029}
  {\bibfield  {journal} {\bibinfo  {journal} {Phys. Rev. Lett.}\ }\textbf
  {\bibinfo {volume} {84}},\ \bibinfo {pages} {3029} (\bibinfo {year}
  {2000})}\BibitemShut {NoStop}%
\bibitem [{\citenamefont {Xun}\ \emph {et~al.}(2020)\citenamefont {Xun},
  \citenamefont {Li}, \citenamefont {Jiao}, \citenamefont {Li},\ and\
  \citenamefont {Tang}}]{Xun_2019}%
  \BibitemOpen
  \bibfield  {author} {\bibinfo {author} {\bibfnamefont {Z.-P.}\ \bibnamefont
  {Xun}}, \bibinfo {author} {\bibfnamefont {J.-X.}\ \bibnamefont {Li}},
  \bibinfo {author} {\bibfnamefont {Y.}~\bibnamefont {Jiao}}, \bibinfo {author}
  {\bibfnamefont {R.-T.}\ \bibnamefont {Li}}, \ and\ \bibinfo {author}
  {\bibfnamefont {G.}~\bibnamefont {Tang}},\ }\href {\doibase
  https://doi.org/10.1016/j.physa.2019.122998} {\bibfield  {journal} {\bibinfo
  {journal} {Physica A}\ }\textbf {\bibinfo {volume} {540}},\ \bibinfo {pages}
  {122998} (\bibinfo {year} {2020})}\BibitemShut {NoStop}%
\bibitem [{\citenamefont {L\'opez}(1999)}]{Lopez_1999}%
  \BibitemOpen
  \bibfield  {author} {\bibinfo {author} {\bibfnamefont {J.~M.}\ \bibnamefont
  {L\'opez}},\ }\href {https://link.aps.org/doi/10.1103/PhysRevLett.83.4594}
  {\bibfield  {journal} {\bibinfo  {journal} {Phys. Rev. Lett.}\ }\textbf
  {\bibinfo {volume} {83}},\ \bibinfo {pages} {4594} (\bibinfo {year}
  {1999})}\BibitemShut {NoStop}%
\bibitem [{\citenamefont {Ramasco}\ \emph {et~al.}(2000)\citenamefont
  {Ramasco}, \citenamefont {L\'opez},\ and\ \citenamefont
  {Rodr\'{\i}guez}}]{Lopez_2000}%
  \BibitemOpen
  \bibfield  {author} {\bibinfo {author} {\bibfnamefont {J.~J.}\ \bibnamefont
  {Ramasco}}, \bibinfo {author} {\bibfnamefont {J.~M.}\ \bibnamefont
  {L\'opez}}, \ and\ \bibinfo {author} {\bibfnamefont {M.~A.}\ \bibnamefont
  {Rodr\'{\i}guez}},\ }\href
  {https://link.aps.org/doi/10.1103/PhysRevLett.84.2199} {\bibfield  {journal}
  {\bibinfo  {journal} {Phys. Rev. Lett.}\ }\textbf {\bibinfo {volume} {84}},\
  \bibinfo {pages} {2199} (\bibinfo {year} {2000})}\BibitemShut {NoStop}%
\bibitem [{\citenamefont {L\'opez}\ and\ \citenamefont
  {Rodr\'{\i}guez}(1996)}]{Lopez_1996}%
  \BibitemOpen
  \bibfield  {author} {\bibinfo {author} {\bibfnamefont {J.~M.}\ \bibnamefont
  {L\'opez}}\ and\ \bibinfo {author} {\bibfnamefont {M.~A.}\ \bibnamefont
  {Rodr\'{\i}guez}},\ }\href
  {https://link.aps.org/doi/10.1103/PhysRevE.54.R2189} {\bibfield  {journal}
  {\bibinfo  {journal} {Phys. Rev. E}\ }\textbf {\bibinfo {volume} {54}},\
  \bibinfo {pages} {R2189} (\bibinfo {year} {1996})}\BibitemShut {NoStop}%
\bibitem [{\citenamefont {López}\ \emph {et~al.}(1997)\citenamefont {López},
  \citenamefont {Rodríguez},\ and\ \citenamefont {Cuerno}}]{Lopez_1997}%
  \BibitemOpen
  \bibfield  {author} {\bibinfo {author} {\bibfnamefont {J.~M.}\ \bibnamefont
  {López}}, \bibinfo {author} {\bibfnamefont {M.~A.}\ \bibnamefont
  {Rodríguez}}, \ and\ \bibinfo {author} {\bibfnamefont {R.}~\bibnamefont
  {Cuerno}},\ }\href {\doibase https://doi.org/10.1016/S0378-4371(97)00375-0}
  {\bibfield  {journal} {\bibinfo  {journal} {Physica A}\ }\textbf {\bibinfo
  {volume} {246}},\ \bibinfo {pages} {329} (\bibinfo {year}
  {1997})}\BibitemShut {NoStop}%
\bibitem [{\citenamefont {L\'opez}\ \emph {et~al.}(2005)\citenamefont
  {L\'opez}, \citenamefont {Castro},\ and\ \citenamefont
  {Gallego}}]{Lopez_2005}%
  \BibitemOpen
  \bibfield  {author} {\bibinfo {author} {\bibfnamefont {J.~M.}\ \bibnamefont
  {L\'opez}}, \bibinfo {author} {\bibfnamefont {M.}~\bibnamefont {Castro}}, \
  and\ \bibinfo {author} {\bibfnamefont {R.}~\bibnamefont {Gallego}},\ }\href
  {https://link.aps.org/doi/10.1103/PhysRevLett.94.166103} {\bibfield
  {journal} {\bibinfo  {journal} {Phys. Rev. Lett.}\ }\textbf {\bibinfo
  {volume} {94}},\ \bibinfo {pages} {166103} (\bibinfo {year}
  {2005})}\BibitemShut {NoStop}%
\bibitem [{\citenamefont {Szendro}\ \emph {et~al.}(2007)\citenamefont
  {Szendro}, \citenamefont {L\'opez},\ and\ \citenamefont
  {Rodr\'{\i}guez}}]{Szendro_2007}%
  \BibitemOpen
  \bibfield  {author} {\bibinfo {author} {\bibfnamefont {I.~G.}\ \bibnamefont
  {Szendro}}, \bibinfo {author} {\bibfnamefont {J.~M.}\ \bibnamefont
  {L\'opez}}, \ and\ \bibinfo {author} {\bibfnamefont {M.~A.}\ \bibnamefont
  {Rodr\'{\i}guez}},\ }\href
  {https://link.aps.org/doi/10.1103/PhysRevE.76.011603} {\bibfield  {journal}
  {\bibinfo  {journal} {Phys. Rev. E}\ }\textbf {\bibinfo {volume} {76}},\
  \bibinfo {pages} {011603} (\bibinfo {year} {2007})}\BibitemShut {NoStop}%
\bibitem [{\citenamefont {Al\'es}\ and\ \citenamefont
  {L\'opez}(2019)}]{Lopez_2019}%
  \BibitemOpen
  \bibfield  {author} {\bibinfo {author} {\bibfnamefont {A.}~\bibnamefont
  {Al\'es}}\ and\ \bibinfo {author} {\bibfnamefont {J.~M.}\ \bibnamefont
  {L\'opez}},\ }\href {https://link.aps.org/doi/10.1103/PhysRevE.99.062139}
  {\bibfield  {journal} {\bibinfo  {journal} {Phys. Rev. E}\ }\textbf {\bibinfo
  {volume} {99}},\ \bibinfo {pages} {062139} (\bibinfo {year}
  {2019})}\BibitemShut {NoStop}%
\bibitem [{\citenamefont {Alés}\ and\ \citenamefont
  {López}(2020)}]{Lopez_2020}%
  \BibitemOpen
  \bibfield  {author} {\bibinfo {author} {\bibfnamefont {A.}~\bibnamefont
  {Alés}}\ and\ \bibinfo {author} {\bibfnamefont {J.~M.}\ \bibnamefont
  {López}},\ }\href {\doibase 10.1088/1742-5468/ab74c9} {\bibfield  {journal}
  {\bibinfo  {journal} {J. Stat. Mech.: Theory Exp.}\ }\textbf {\bibinfo
  {volume} {2020}},\ \bibinfo {pages} {033210} (\bibinfo {year}
  {2020})}\BibitemShut {NoStop}%
\bibitem [{\citenamefont {Al\'es}\ and\ \citenamefont
  {L\'opez}(2021)}]{Lopez_2021}%
  \BibitemOpen
  \bibfield  {author} {\bibinfo {author} {\bibfnamefont {A.}~\bibnamefont
  {Al\'es}}\ and\ \bibinfo {author} {\bibfnamefont {J.~M.}\ \bibnamefont
  {L\'opez}},\ }\href {https://link.aps.org/doi/10.1103/PhysRevE.104.044108}
  {\bibfield  {journal} {\bibinfo  {journal} {Phys. Rev. E}\ }\textbf {\bibinfo
  {volume} {104}},\ \bibinfo {pages} {044108} (\bibinfo {year}
  {2021})}\BibitemShut {NoStop}%
\bibitem [{\citenamefont {Purrello}\ \emph {et~al.}(2019)\citenamefont
  {Purrello}, \citenamefont {Iguain},\ and\ \citenamefont
  {Kolton}}]{Kolton_2019}%
  \BibitemOpen
  \bibfield  {author} {\bibinfo {author} {\bibfnamefont {V.~H.}\ \bibnamefont
  {Purrello}}, \bibinfo {author} {\bibfnamefont {J.~L.}\ \bibnamefont
  {Iguain}}, \ and\ \bibinfo {author} {\bibfnamefont {A.~B.}\ \bibnamefont
  {Kolton}},\ }\href {https://link.aps.org/doi/10.1103/PhysRevE.99.032105}
  {\bibfield  {journal} {\bibinfo  {journal} {Phys. Rev. E}\ }\textbf {\bibinfo
  {volume} {99}},\ \bibinfo {pages} {032105} (\bibinfo {year}
  {2019})}\BibitemShut {NoStop}%
\bibitem [{\citenamefont {Sneppen}(1992)}]{Sneppen_1992}%
  \BibitemOpen
  \bibfield  {author} {\bibinfo {author} {\bibfnamefont {K.}~\bibnamefont
  {Sneppen}},\ }\href {https://link.aps.org/doi/10.1103/PhysRevLett.69.3539}
  {\bibfield  {journal} {\bibinfo  {journal} {Phys. Rev. Lett.}\ }\textbf
  {\bibinfo {volume} {69}},\ \bibinfo {pages} {3539} (\bibinfo {year}
  {1992})}\BibitemShut {NoStop}%
\bibitem [{\citenamefont {Maslov}\ and\ \citenamefont
  {Zhang}(1995)}]{Maslov_1995}%
  \BibitemOpen
  \bibfield  {author} {\bibinfo {author} {\bibfnamefont {S.}~\bibnamefont
  {Maslov}}\ and\ \bibinfo {author} {\bibfnamefont {Y.-C.}\ \bibnamefont
  {Zhang}},\ }\href {https://link.aps.org/doi/10.1103/PhysRevLett.75.1550}
  {\bibfield  {journal} {\bibinfo  {journal} {Phys. Rev. Lett.}\ }\textbf
  {\bibinfo {volume} {75}},\ \bibinfo {pages} {1550} (\bibinfo {year}
  {1995})}\BibitemShut {NoStop}%
\bibitem [{\citenamefont {Rodr\'{\i}guez-Fern\'andez}\ \emph
  {et~al.}(2022)\citenamefont {Rodr\'{\i}guez-Fern\'andez}, \citenamefont
  {Santalla}, \citenamefont {Castro},\ and\ \citenamefont
  {Cuerno}}]{Cuerno_2022}%
  \BibitemOpen
  \bibfield  {author} {\bibinfo {author} {\bibfnamefont {E.}~\bibnamefont
  {Rodr\'{\i}guez-Fern\'andez}}, \bibinfo {author} {\bibfnamefont {S.~N.}\
  \bibnamefont {Santalla}}, \bibinfo {author} {\bibfnamefont {M.}~\bibnamefont
  {Castro}}, \ and\ \bibinfo {author} {\bibfnamefont {R.}~\bibnamefont
  {Cuerno}},\ }\href {https://link.aps.org/doi/10.1103/PhysRevE.106.024802}
  {\bibfield  {journal} {\bibinfo  {journal} {Phys. Rev. E}\ }\textbf {\bibinfo
  {volume} {106}},\ \bibinfo {pages} {024802} (\bibinfo {year}
  {2022})}\BibitemShut {NoStop}%
\bibitem [{\citenamefont {Zaitsev}(1992)}]{Zaitsev_1992}%
  \BibitemOpen
  \bibfield  {author} {\bibinfo {author} {\bibfnamefont {S.}~\bibnamefont
  {Zaitsev}},\ }\href {\doibase https://doi.org/10.1016/0378-4371(92)90053-S}
  {\bibfield  {journal} {\bibinfo  {journal} {Physica A}\ }\textbf {\bibinfo
  {volume} {189}},\ \bibinfo {pages} {411} (\bibinfo {year}
  {1992})}\BibitemShut {NoStop}%
\bibitem [{\citenamefont {C\'ordoba-Torres}\ \emph {et~al.}(2009)\citenamefont
  {C\'ordoba-Torres}, \citenamefont {Mesquita}, \citenamefont {Bastos},\ and\
  \citenamefont {Nogueira}}]{Bastos_2009}%
  \BibitemOpen
  \bibfield  {author} {\bibinfo {author} {\bibfnamefont {P.}~\bibnamefont
  {C\'ordoba-Torres}}, \bibinfo {author} {\bibfnamefont {T.~J.}\ \bibnamefont
  {Mesquita}}, \bibinfo {author} {\bibfnamefont {I.~N.}\ \bibnamefont
  {Bastos}}, \ and\ \bibinfo {author} {\bibfnamefont {R.~P.}\ \bibnamefont
  {Nogueira}},\ }\href
  {https://link.aps.org/doi/10.1103/PhysRevLett.102.055504} {\bibfield
  {journal} {\bibinfo  {journal} {Phys. Rev. Lett.}\ }\textbf {\bibinfo
  {volume} {102}},\ \bibinfo {pages} {055504} (\bibinfo {year}
  {2009})}\BibitemShut {NoStop}%
\end{thebibliography}%






\end{document}